\documentclass[showpacs,twocolumn,superscriptaddress]{revtex4}

\usepackage{graphicx}
\usepackage{dcolumn}
\usepackage{bm}
\usepackage{amsmath}

\begin{document}
\title{Single-parameter quantized charge pumping in high magnetic fields}

\author{B. Kaestner}
\email[Electronic address: ]{Bernd.Kaestner@ptb.de}
\author{Ch. Leicht}
\affiliation{
		Physikalisch-Technische Bundesanstalt, Bundesallee 100, 38116 Braunschweig, Germany
}

\author{V. Kashcheyevs}
\affiliation{
		Institute for Solid State Physics, University of Latvia, Riga LV-1063, Latvia
}
\affiliation{Faculty of Physics and Mathematics,
                  University of Latvia, Ze{\c{l}\c{l}}u
                  street 8, Riga LV-1002, Latvia}

\author{K. Pierz$^1$}
\author{U. Siegner$^1$}
\author{H. W. Schumacher$^1$}

\date{\today}

\begin{abstract}

We study single-parameter quantized charge pumping via a semiconductor quantum dot in high magnetic fields. The quantum dot is defined between two top gates in an AlGaAs/GaAs heterostructure. Application of an oscillating voltage to one of the gates leads to pumped current plateaus in the gate characteristic, corresponding to controlled transfer of integer multiples of electrons per cycle. In a perpendicular-to-plane magnetic field the plateaus become more pronounced indicating an improved current quantization. Current quantization is sustained up to magnetic fields where full spin polarization of the device can be expected.

\end{abstract}


\maketitle

Generating well defined currents by manipulating single charges has attracted  considerable interest in the last two decades from both fundamental and applied points of view~\cite{Averin1991}. A particular potential of application lies in the field of metrology to provide a link between time and current units~\cite{mills2006}. Different approaches have been studied, such as arrays of tunnel-connected metallic islands controlled by a number of phase shifted ac signals~\cite{geerligs1990, pothier1PBI, Keller1996, pekola2007} or semiconducting channels along which the potential can be modulated continuously~\cite{Kouwenhoven1PBI, Shilton1996, fujiwara2008, blumenthal2007a, Kaestner2007c}. The pumping mechanism demonstrated in Ref.~\cite{blumenthal2007a} allows gigahertz pumping comparable to surface-acoustic-wave pumps~\cite{Shilton1996} while promising a higher degree of control. It employs three electrodes of which two are modulated at a fixed phase shift and with different amplitudes. In Ref.~\cite{Kaestner2007c} it was shown that a single modulated voltage signal is sufficient to operate the pump and a numerical investigation indicated the importance of the tunnel barrier shape for improving the accuracy. A possible way to manipulate these tunnel couplings might be the application of a magnetic field owing to its influence on the wave function and the corresponding rearrangement of electrons between quantum states (see for instance~\cite{ashoori1PBI}). 
Therefore the operation of such a single-parameter charge pump has been realized in the presented work when a perpendicular-to-plane magnetic field was applied.

\begin{figure}
\includegraphics{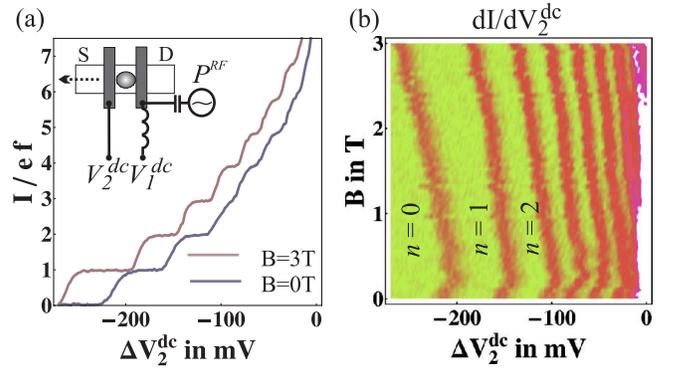}
\caption{\label{fig:bSweep} (Color online) (a) Pumped current $I$ normalized by $e f$ versus the variation of the dc voltage applied to gate 2, $\Delta V_2^{dc}$. The inset shows a schematic of the device, with the dashed arrow indicating the direction of the pumped electrons. S and D indicate source and drain contacts. (b) $dI/dV_2^{dc}$ in arbitrary units as a function of applied $B$-field and $\Delta V_2^{dc}$. The number of electrons, $n$, pumped per cycle is specified in the figure. The measurements where performed in device 1 at $f = 50\,$MHz, base temperature $T = 40\,$mK and power $P^{RF} = -16\,$dBm.}
\end{figure}

Two devices have been investigated which were both realized in an AlGaAs/GaAs heterostructure with a carrier concentration of $2.1 \times 10^{15}\,$m$^{-2}$ and a mobility of $97\,$m$^2$/Vs in the dark. A $700\,$nm wide wire connected to two-dimensional electron gases was created by wet-etching the doped AlGaAs layer. 
This channel is crossed by two $100\,$nm wide Ti-Au finger gates of 250$\,$nm separation. A schematic is shown in the inset of Fig.~\ref{fig:bSweep}(a). A quantum dot (QD) with discrete quasibound states between the gates can be created by applying sufficiently large negative voltages $V_1$ and $V_2$ to gate 1 and gate 2, respectively. An additional radiofrequency (rf) signal is coupled to gate 1. The resulting voltages are therefore $V_1 = V^{dc}_1 + V^{ac}_1 \cos(2 \pi f t)$ and $V_2 = V^{dc}_2$ at gate 1 and gate 2, respectively. If the oscillation amplitude is high enough, then the bound state drops below the Fermi level during the first half-cycle of the periodic signal and can be loaded with electrons from source. During the second half-cycle the bound state is raised sufficiently fast to avoid backtunneling and electrons can be unloaded to the drain. The resulting current changes in steps of $e f$ as the voltage parameters are varied, where $e$ is the electron charge and $f$ is the pumping frequency. For a detailed discussion we refer to Ref.~\cite{Kaestner2007c}.

The measured characteristic for device 1 is shown in Fig.~\ref{fig:bSweep}. The measurements were performed at a base temperature of $T = 40\,$mK. The pumping frequency was chosen to be $f = 50\,$MHz at a rf power of $P^{RF} = -16\,$dBm. As shown in Fig.~\ref{fig:bSweep}(a) the pumped current increases in steps of $e f$ as $V_2$ is made more positive. The red and blue curve compare the dependence of the pumped dc current on $V_2$ without and with perpendicular magnetic field of $B=3\,$T applied, respectively. With the applied field the step-edges shift and become steeper.  The evolution of the step-edges can be seen in more detail in Fig.~\ref{fig:bSweep}(b), showing the derivative $dI/dV_2$ in arbitrary units as a function of $V_2$ and $B$. Below $0.3\,$T the step-edges first shift towards more positive voltages. Then for fields between 0.3 and $3\,$T a linear shift towards more negative voltages is found. The shift towards more negative voltage may be linked to the increase in confinement of the charges, assuming the current drop at the step-edge is caused by backtunneling of electrons loaded during the first half-cycle.

\begin{figure}
\includegraphics{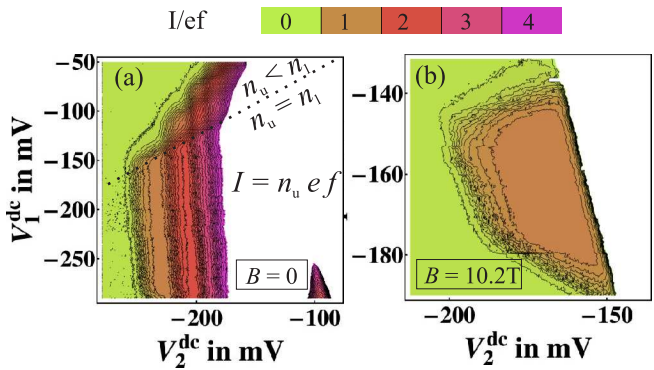}
\caption{\label{fig:2DChar} (Color online) Pumped current at $f=500\,$MHz through device 2 as a function of $V_1^{dc}$ and $V_2^{dc}$. In (a) no magnetic field was applied. The dotted line indicates the transition above which the emission of electrons to the drain, $n_u$, is not complete and smaller than the number of loaded electrons, $n_l$. In (b) the pumped current is shown when a perpendicular-to-plane field of $B=10.2\,$T was applied.}
\end{figure}

For intermediate fields above $3\,$T the plateau structure of the pumped current evolves in a more complex way, which will be subject to future investigations. Now we focus on the behaviour in high magnetic field.
An example for quantized pumping in the high field regime is shown in Fig.~\ref{fig:2DChar} for a second device. Here, the dependence of the pumped current on the two gate voltage parameters $V_1^{dc}$ and $V_2^{dc}$ is shown for $B=0$ (a) and $B=10.2\,$T (b), respectively. Note that in this device pumping occurs at a different set of voltage parameters due to different effective dimensions of the QD, leads, tunneling barriers and according changes in the electronic properties. Device 2 was operated at a higher pumping frequency of $f = 500\,$MHz and higher base temperature of $T=300\,$mK. The rf power was set to $P^{rf} = -23\,$dBm.
In Fig.~\ref{fig:2DChar}(a) up to four plateaus corresponding to quantized currents of $I=n e f$ ($n = 1,2,3,4$) are observed.
The tilted dashed line indicates the border above which the number of electrons, $n_l$, loaded during the first half-cycle is larger than the number of electrons $n_u$, which have sufficiently high energy to be unloaded to the drain reservoir~\cite{kaestner2008} during the second half-cycle. Any electrons remaining in the dot will be deposited back to the source. The condition $n_u = n_l$ typically leads to a higher accuracy and robustness~\cite{kaestner2008}. 

\begin{figure}
\includegraphics{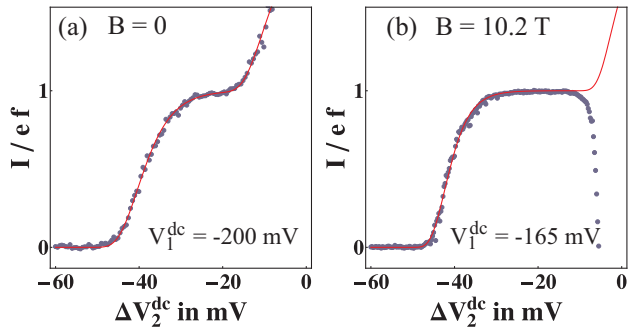}
\caption{\label{fig:1DChar} (Color online) Pumped current $I$ normalized by $e f$ at $f=500\,$MHz through device 2 versus the variation of the dc voltage applied to gate 2, $\Delta V_2^{dc}$, at fixed $V_1^{dc}$. In (a) no field was applied, while (b) shows the characteristic for $B = 10.2\,$T. The gate characteristic was fitted using Eq.~\ref{eq:fit} together with the following parameters: For zero field (a), $V_a = -40.4\,$mV, $V_b = 4.6\,$mV, $\Delta_2 = 6.4$ while for the measurements at $B = 10.2\,$ (b) the values $V_a = -42.32\,$mV, $V_b = 3.31\,$mV, and $\Delta_2 = 12$ were used.}
\end{figure}

Fig.~\ref{fig:2DChar}(b) shows the gate voltage dependence of the pumped current at $B=10.2\,$T. Only one plateau corresponding to $I = e f$ is observed and multi-electron plateaus are completely suppressed. At the same time the single-electron plateau is extended along $V^{dc}_2$. This is shown more clearly in Fig~\ref{fig:1DChar} for a particular voltage $V^{dc}_1$. Part (a) shows the plateau region without magnetic field. The plateaus in zero field are not as well pronounced as in device 1. However, the quantization improves significantly when a magnetic field is applied, as shown in part (b).
In the latter case, as $V^{dc}_2$ is tuned to more positive values the current drops sharply and becomes negative, whereas without magnetic field the current continues to increase, showing further plateaus near multiples of $e f$. The sharp current drop as $V_2^{dc}$ becomes more and more positive might be due to increasing tunneling coupling to the drain reservoir such that a large amount of charges enters and leaves the dot through drain. The resulting $ac$ current might be rectified, for instance by nonlinearities of the amplifier, eliminating the quantization plateau and leading to the measured negative $dc$ current. Note that this sharp transition to large and unquantized current also occurs at zero field but at more positive values of $V_2^{dc}$.

We now evaluate the plateau shape in more detail. For quantization steps dominated by backtunneling, the occupation probability $P(t)$ of the dot switches from $1$ at most positive $V_1$ to some well-defined $P_0$ as the charge state is energetically lifted up from the Fermi sea. Denoting by $t_0$ the starting moment of backtunneling, and observing~\cite{Kaestner2007c} that its rate depends exponentially both on  $V_{1}$ (via entrance barrier height) and $V_2$ (via level position),
we can describe the time evolution of $P(t)$ as [see Eq.(2) of~\cite{Kaestner2007c}]
\begin{align} \label{eq:rateeqsimple}
  \dot{P}(t) & = - \Gamma_L(t) P(t) \text{ with  } P(t_0) =1  \\
  \Gamma_{L}(t) & \equiv \Gamma_0[ V_1(t_0) ] \exp [- \alpha V_2^{dc} - \beta  t ],
  \label{eq:gammasimple}
\end{align}
where $\alpha >0$, and $\beta \propto -d V_1/d t \propto f V_1^{ac} > 0$ are constants and $\Gamma_0$ is the starting rate for backtunneling. Eq.~\eqref{eq:rateeqsimple} holds for a limited fraction of the period when  $V_2^{dc}$ is negative enough to make the exit barrier blocked.
Estimating $P_0$ from the limit $t\to \infty$ of the solution to Eq.~\eqref{eq:rateeqsimple} gives $P_0 = \exp [-(\Gamma_0/\beta) \exp (-\alpha V_2^{dc}) ]$.

A phenomenological generalization of this theory to  the case of several steps is  to choose individual starting rates  $\Gamma_0 \propto \exp \Delta_i$ for each of the backtunneling transitions $i \to i-1$ ($i=1,2 \ldots n$) and sum up the corresponding values of the transferred charge per period $e P_0$. A fitting formula corresponding to this procedure is
\begin{equation}
 I/ef = \sum_{i=1}^n \exp(-\exp(-\frac{V^{dc}_2-V_a}{V_b}+\Delta_i)),
 \label{eq:fit}
\end{equation}
where $V_a, V_b$ and $\Delta_i$ will be used to fit the data points. The data in Fig.~\ref{fig:1DChar} are modeled by including the first two plateaus, i.e. $n=2$. Parameters $V_a$ and $V_b$ are chosen so that always $\Delta_1 = 0$. In this way, $\Delta_2$ determines the flatness of the first plateau independently of position and scale of the plateau structure along the voltage axis. The parameters for modeling the data are given in Fig.~\ref{fig:1DChar}. In particular, $\Delta_2 (B=0) = 6.4$ for pumping in zero field, while $\Delta_2 (B=10.2\,T) = 12$ and is significantly increased when the field was applied. $\Delta_2(B=10.2\,T)$ was chosen under the assumption that the second plateau would start just outside the region where pumping can be achieved. Operating the pump exactly in the middle of the $e f$ plateau this model predicts a deviation from the exact value of $1-\exp(-\exp(-\Delta_2/2))$, so that $B$-field operation has reduced the error by an order of magnitude from $4\,$\% to $0.3\,$\%. The physical meaning of $\Delta_2$ may be related to how the functional dependence of $\Gamma_0(V_1)$ varies between the different charging states.

Note that for the investigated devices full spin polarization of source and drain (filling factor one) occurs at an applied field of $8.9\,$T. Therefore at the higher applied field of $10.2\,$T only spin polarized electrons are loaded from the source. As the spin flip times of the electron on the quantum dot should be significantly longer than the average hold time during pumping~\cite{elzerman2004} fully spin polarized quantized charge pumping through the device can be expected.
Furthermore the spin state of source and drain at the given field could be independently tuned via the carrier density by additional top gates on each side of the QD. Such a device might thus serve as a clocked source of spin polarized single electrons for spin electronics and quantum information processing.

From the investigation above we conclude that the accuracy in single-parameter pumping can be improved by the application of a high perpendicular-to-plane magnetic field~\cite{remarkNPL2008}. The phenomenon of quantized pumping can be sustained in high fields in contrast to the conceptually related charge pumps using surface acoustic waves~\cite{cunningham2}, allowing spin polarized quantized charge pumping. So far we have tested and observed quantized pumping in such devices up to fields of $16\,$T. The magnetic field as an additional control parameter makes single-parameter pumps in AlGaAs/GaAs heterostructures promising candidates for an accurately quantized, large-current source as needed for fundamental experiments in metrology and quantum electronics.

\begin{acknowledgments}
The research conducted within this EURAMET joint research project has received
funding from the European Community's Seventh Framework Programme, ERANET
Plus, under Grant Agreement No. 217257. Assistance with device fabrication from Th. Weimann, P. Hinze and H. Marx and helpful discussions with L. Schweitzer are greatly acknowledged. V.K. has been supported by the Latvian Council of Science.
\end{acknowledgments}


\begin{thebibliography}{16}
\expandafter\ifx\csname natexlab\endcsname\relax\def\natexlab#1{#1}\fi
\expandafter\ifx\csname bibnamefont\endcsname\relax
  \def\bibnamefont#1{#1}\fi
\expandafter\ifx\csname bibfnamefont\endcsname\relax
  \def\bibfnamefont#1{#1}\fi
\expandafter\ifx\csname citenamefont\endcsname\relax
  \def\citenamefont#1{#1}\fi
\expandafter\ifx\csname url\endcsname\relax
  \def\url#1{\texttt{#1}}\fi
\expandafter\ifx\csname urlprefix\endcsname\relax\def\urlprefix{URL }\fi
\providecommand{\bibinfo}[2]{#2}
\providecommand{\eprint}[2][]{\url{#2}}

\bibitem[{\citenamefont{Averin and Likharev}(1991)}]{Averin1991}
\bibinfo{author}{\bibfnamefont{D.~V.} \bibnamefont{Averin}} \bibnamefont{and}
  \bibinfo{author}{\bibfnamefont{K.~K.} \bibnamefont{Likharev}},
  \emph{\bibinfo{title}{Mesoscopic Phenomena in Solids}}
  (\bibinfo{publisher}{Elsevier, Amsterdam}, \bibinfo{year}{1991}), pp.
  \bibinfo{pages}{173 -- 271}.

\bibitem[{\citenamefont{Mills et~al.}(2006)\citenamefont{Mills, Mohr, Quinn,
  Taylor, and Williams}}]{mills2006}
\bibinfo{author}{\bibfnamefont{I.~M.} \bibnamefont{Mills}},
  \bibinfo{author}{\bibfnamefont{P.~J.} \bibnamefont{Mohr}},
  \bibinfo{author}{\bibfnamefont{T.~J.} \bibnamefont{Quinn}},
  \bibinfo{author}{\bibfnamefont{B.~N.} \bibnamefont{Taylor}},
  \bibnamefont{and} \bibinfo{author}{\bibfnamefont{E.~R.}
  \bibnamefont{Williams}}, \bibinfo{journal}{Metrologia}
  \textbf{\bibinfo{volume}{43}}, \bibinfo{pages}{227} (\bibinfo{year}{2006}).

\bibitem[{\citenamefont{Geerligs et~al.}(1990)\citenamefont{Geerligs, Anderegg,
  Holweg, Mooij, Pothier, Esteve, Urbina, and Devoret}}]{geerligs1990}
\bibinfo{author}{\bibfnamefont{L.~J.} \bibnamefont{Geerligs}},
  \bibinfo{author}{\bibfnamefont{V.~F.} \bibnamefont{Anderegg}},
  \bibinfo{author}{\bibfnamefont{P.~A.~M.} \bibnamefont{Holweg}},
  \bibinfo{author}{\bibfnamefont{J.~E.} \bibnamefont{Mooij}},
  \bibinfo{author}{\bibfnamefont{H.}~\bibnamefont{Pothier}},
  \bibinfo{author}{\bibfnamefont{D.}~\bibnamefont{Esteve}},
  \bibinfo{author}{\bibfnamefont{C.}~\bibnamefont{Urbina}}, \bibnamefont{and}
  \bibinfo{author}{\bibfnamefont{M.~H.} \bibnamefont{Devoret}},
  \bibinfo{journal}{Phys. Rev. Let.} \textbf{\bibinfo{volume}{64}},
  \bibinfo{pages}{2691} (\bibinfo{year}{1990}).

\bibitem[{\citenamefont{Pothier et~al.}(1992)\citenamefont{Pothier, Lafarge,
  Urbina, Esteve, and Devoret}}]{pothier1PBI}
\bibinfo{author}{\bibfnamefont{H.}~\bibnamefont{Pothier}},
  \bibinfo{author}{\bibfnamefont{P.}~\bibnamefont{Lafarge}},
  \bibinfo{author}{\bibfnamefont{C.}~\bibnamefont{Urbina}},
  \bibinfo{author}{\bibfnamefont{D.}~\bibnamefont{Esteve}}, \bibnamefont{and}
  \bibinfo{author}{\bibfnamefont{M.~H.} \bibnamefont{Devoret}},
  \bibinfo{journal}{Europhys. Lett.} \textbf{\bibinfo{volume}{17}},
  \bibinfo{pages}{249} (\bibinfo{year}{1992}).

\bibitem[{\citenamefont{Keller et~al.}(1996)\citenamefont{Keller, Martinis,
  Zimmerman, and Steinbach}}]{Keller1996}
\bibinfo{author}{\bibfnamefont{M.~W.} \bibnamefont{Keller}},
  \bibinfo{author}{\bibfnamefont{J.~M.} \bibnamefont{Martinis}},
  \bibinfo{author}{\bibfnamefont{N.~M.} \bibnamefont{Zimmerman}},
  \bibnamefont{and} \bibinfo{author}{\bibfnamefont{A.~H.}
  \bibnamefont{Steinbach}}, \bibinfo{journal}{Appl. Phys. Lett.}
  \textbf{\bibinfo{volume}{69}}, \bibinfo{pages}{1804} (\bibinfo{year}{1996}).

\bibitem[{\citenamefont{Pekola et~al.}(2008)\citenamefont{Pekola, Vartiainen,
  M{\"o}tt{\"o}nen, Saira, Meschke, and Averin}}]{pekola2007}
\bibinfo{author}{\bibfnamefont{J.~P.} \bibnamefont{Pekola}},
  \bibinfo{author}{\bibfnamefont{J.~J.} \bibnamefont{Vartiainen}},
  \bibinfo{author}{\bibfnamefont{M.}~\bibnamefont{M{\"o}tt{\"o}nen}},
  \bibinfo{author}{\bibfnamefont{O.-P.} \bibnamefont{Saira}},
  \bibinfo{author}{\bibfnamefont{M.}~\bibnamefont{Meschke}}, \bibnamefont{and}
  \bibinfo{author}{\bibfnamefont{D.~V.} \bibnamefont{Averin}},
  \bibinfo{journal}{Nature Physics} \textbf{\bibinfo{volume}{4}},
  \bibinfo{pages}{120} (\bibinfo{year}{2008}).

\bibitem[{\citenamefont{Kouwenhoven et~al.}(1991)\citenamefont{Kouwenhoven,
  Johnson, {van der Vaart}, Harmans, and Foxon}}]{Kouwenhoven1PBI}
\bibinfo{author}{\bibfnamefont{L.~P.} \bibnamefont{Kouwenhoven}},
  \bibinfo{author}{\bibfnamefont{A.~T.} \bibnamefont{Johnson}},
  \bibinfo{author}{\bibfnamefont{N.~C.} \bibnamefont{{van der Vaart}}},
  \bibinfo{author}{\bibfnamefont{C.~J. P.~M.} \bibnamefont{Harmans}},
  \bibnamefont{and} \bibinfo{author}{\bibfnamefont{C.~T.} \bibnamefont{Foxon}},
  \bibinfo{journal}{Phys. Rev. Lett.} \textbf{\bibinfo{volume}{67}},
  \bibinfo{pages}{1626} (\bibinfo{year}{1991}).

\bibitem[{\citenamefont{Shilton et~al.}(1996)\citenamefont{Shilton, Talyanskii,
  Pepper, Ritchie, Frost, Ford, Smith, and Jones}}]{Shilton1996}
\bibinfo{author}{\bibfnamefont{J.~M.} \bibnamefont{Shilton}},
  \bibinfo{author}{\bibfnamefont{V.~I.} \bibnamefont{Talyanskii}},
  \bibinfo{author}{\bibfnamefont{M.}~\bibnamefont{Pepper}},
  \bibinfo{author}{\bibfnamefont{D.~A.} \bibnamefont{Ritchie}},
  \bibinfo{author}{\bibfnamefont{J.~E.~F.} \bibnamefont{Frost}},
  \bibinfo{author}{\bibfnamefont{C.~J.~B.} \bibnamefont{Ford}},
  \bibinfo{author}{\bibfnamefont{C.~G.} \bibnamefont{Smith}}, \bibnamefont{and}
  \bibinfo{author}{\bibfnamefont{G.~A.~C.} \bibnamefont{Jones}},
  \bibinfo{journal}{J. Phys.: Condens. Matter} \textbf{\bibinfo{volume}{8}},
  \bibinfo{pages}{L531} (\bibinfo{year}{1996}).

\bibitem[{\citenamefont{Fujiwara et~al.}(2008)\citenamefont{Fujiwara,
  Nishiguchi, and Ono}}]{fujiwara2008}
\bibinfo{author}{\bibfnamefont{A.}~\bibnamefont{Fujiwara}},
  \bibinfo{author}{\bibfnamefont{K.}~\bibnamefont{Nishiguchi}},
  \bibnamefont{and} \bibinfo{author}{\bibfnamefont{Y.}~\bibnamefont{Ono}},
  \bibinfo{journal}{Appl. Phys. Lett.} \textbf{\bibinfo{volume}{92}},
  \bibinfo{pages}{042102} (\bibinfo{year}{2008}).

\bibitem[{\citenamefont{Blumenthal et~al.}(2007)\citenamefont{Blumenthal,
  Kaestner, Li, Giblin, Janssen, Pepper, Anderson, Jones, and
  Ritchie}}]{blumenthal2007a}
\bibinfo{author}{\bibfnamefont{M.~D.} \bibnamefont{Blumenthal}},
  \bibinfo{author}{\bibfnamefont{B.}~\bibnamefont{Kaestner}},
  \bibinfo{author}{\bibfnamefont{L.}~\bibnamefont{Li}},
  \bibinfo{author}{\bibfnamefont{S.}~\bibnamefont{Giblin}},
  \bibinfo{author}{\bibfnamefont{T.~J. B.~M.} \bibnamefont{Janssen}},
  \bibinfo{author}{\bibfnamefont{M.}~\bibnamefont{Pepper}},
  \bibinfo{author}{\bibfnamefont{D.}~\bibnamefont{Anderson}},
  \bibinfo{author}{\bibfnamefont{G.}~\bibnamefont{Jones}}, \bibnamefont{and}
  \bibinfo{author}{\bibfnamefont{D.~A.} \bibnamefont{Ritchie}},
  \bibinfo{journal}{Nature Physics} \textbf{\bibinfo{volume}{3}},
  \bibinfo{pages}{343 } (\bibinfo{year}{2007}).

\bibitem[{\citenamefont{Kaestner
  et~al.}(2008{\natexlab{a}})\citenamefont{Kaestner, Kashcheyevs, Amakawa,
  Blumenthal, Li, Janssen, Hein, Pierz, Weimann, Siegner
  et~al.}}]{Kaestner2007c}
\bibinfo{author}{\bibfnamefont{B.}~\bibnamefont{Kaestner}},
  \bibinfo{author}{\bibfnamefont{V.}~\bibnamefont{Kashcheyevs}},
  \bibinfo{author}{\bibfnamefont{S.}~\bibnamefont{Amakawa}},
  \bibinfo{author}{\bibfnamefont{M.~D.} \bibnamefont{Blumenthal}},
  \bibinfo{author}{\bibfnamefont{L.}~\bibnamefont{Li}},
  \bibinfo{author}{\bibfnamefont{T.~J. B.~M.} \bibnamefont{Janssen}},
  \bibinfo{author}{\bibfnamefont{G.}~\bibnamefont{Hein}},
  \bibinfo{author}{\bibfnamefont{K.}~\bibnamefont{Pierz}},
  \bibinfo{author}{\bibfnamefont{T.}~\bibnamefont{Weimann}},
  \bibinfo{author}{\bibfnamefont{U.}~\bibnamefont{Siegner}},
  \bibnamefont{et~al.}, \bibinfo{journal}{Phys. Rev. B}
  \textbf{\bibinfo{volume}{77}}, \bibinfo{pages}{153301}
  (\bibinfo{year}{2008}{\natexlab{a}}), \eprint{arXiv:0707.0993
  [cond-mat]}.

\bibitem[{\citenamefont{Ashoori}(1996)}]{ashoori1PBI}
\bibinfo{author}{\bibfnamefont{R.~C.} \bibnamefont{Ashoori}},
  \bibinfo{journal}{Nature} \textbf{\bibinfo{volume}{379}},
  \bibinfo{pages}{413} (\bibinfo{year}{1996}).

\bibitem[{\citenamefont{Kaestner
  et~al.}(2008{\natexlab{b}})\citenamefont{Kaestner, Kashcheyevs, Hein, Pierz,
  Siegner, and Schumacher}}]{kaestner2008}
\bibinfo{author}{\bibfnamefont{B.}~\bibnamefont{Kaestner}},
  \bibinfo{author}{\bibfnamefont{V.}~\bibnamefont{Kashcheyevs}},
  \bibinfo{author}{\bibfnamefont{G.}~\bibnamefont{Hein}},
  \bibinfo{author}{\bibfnamefont{K.}~\bibnamefont{Pierz}},
  \bibinfo{author}{\bibfnamefont{U.}~\bibnamefont{Siegner}}, \bibnamefont{and}
  \bibinfo{author}{\bibfnamefont{H.~W.} \bibnamefont{Schumacher}},
  \bibinfo{journal}{Appl. Phys. Lett.} \textbf{\bibinfo{volume}{92}},
  \bibinfo{pages}{192106} (\bibinfo{year}{2008}{\natexlab{b}}).

\bibitem[{\citenamefont{Elzerman et~al.}(2004)\citenamefont{Elzerman, Hanson,
  {Willems van Beveren}, Witkamp, Vandersypen, and Kouwenhoven}}]{elzerman2004}
\bibinfo{author}{\bibfnamefont{J.~M.} \bibnamefont{Elzerman}},
  \bibinfo{author}{\bibfnamefont{R.}~\bibnamefont{Hanson}},
  \bibinfo{author}{\bibfnamefont{L.~H.} \bibnamefont{{Willems van Beveren}}},
  \bibinfo{author}{\bibfnamefont{B.}~\bibnamefont{Witkamp}},
  \bibinfo{author}{\bibfnamefont{L.~M.~K.} \bibnamefont{Vandersypen}},
  \bibnamefont{and} \bibinfo{author}{\bibfnamefont{L.~P.}
  \bibnamefont{Kouwenhoven}}, \bibinfo{journal}{Nature}
  \textbf{\bibinfo{volume}{430}}, \bibinfo{pages}{431} (\bibinfo{year}{2004}).

\bibitem[{rem()}]{remarkNPL2008}
\bibinfo{note}{During the preparation of our manuscript we became aware of
  similar findings in the low-field regime by S. J. Wright \emph{et al.}
  arXiv:0811.0494v1 [cond-mat].}

\bibitem[{\citenamefont{Cunningham et~al.}(2000)\citenamefont{Cunningham,
  Talyanskii, Shilton, Pepper, Kristensen, and Lindelof}}]{cunningham2}
\bibinfo{author}{\bibfnamefont{J.}~\bibnamefont{Cunningham}},
  \bibinfo{author}{\bibfnamefont{V.~I.} \bibnamefont{Talyanskii}},
  \bibinfo{author}{\bibfnamefont{J.~M.} \bibnamefont{Shilton}},
  \bibinfo{author}{\bibfnamefont{M.}~\bibnamefont{Pepper}},
  \bibinfo{author}{\bibfnamefont{A.}~\bibnamefont{Kristensen}},
  \bibnamefont{and} \bibinfo{author}{\bibfnamefont{P.~E.}
  \bibnamefont{Lindelof}}, \bibinfo{journal}{Phys. Rev. B}
  \textbf{\bibinfo{volume}{62}}, \bibinfo{pages}{1564} (\bibinfo{year}{2000}).

\end{thebibliography}

\end{document}